\documentclass[EN]{rcftex}
\usepackage{kpfonts}
\usepackage{cite}

\title{Impact of grain properties on the penetration of intruders near a wall into granular matter: a DEM study}

\titleEN{Impacto de las propiedades de los granos en la penetración de intrusos cerca de una frontera en un medio granular: Un estudio DEM} 

\author{M. Espinosa\toaff{a},E . Altshuler\toaff{a\emailto}}

\affiliations{
Group of Complex Systems and Statistical Physics. Physics Faculty, University of Havana, San Lázaro and L, 10400 La Habana, Cuba.  %ap@fisica.cu
\aff Department of General Physics. marcos.espinosa@fisica.uh.cu
\aff Department of Applied Physics. ealtshuler@fisica.uh.cu$^\emailto$
}

\keyW{granular system (sistema granular) 45.70.-n, discrete element method (método de elementos discretos) 45.70.Mg, numerical simulations (simulaciones numéricas) 79.20.Ap}

\begin{document}

\maketitle

Due to the nature of inter-grain interactions --especially their dissipative character-- granular materials show many unexpected phenomena \cite{le1996ticking,eggers1999sand,shinbrot2004granular,martinez2007uphill,altshuler2008revolving,sanchez2014note,altshuler2014settling}. However, analytical tools are often a difficult option to explain them, due to their complexity and the large number of degrees of freedom involved. The discrete element method (DEM) \cite{cundall1979discrete} has proven to be an excellent possibility for the purpose \cite{tanaka2002discrete,gu2014investigation,shi2015investigation,rathbone2015accurate}. Here we use DEM simulations to study the sensitivity of a granular system to the change of the mechanical properties of the grains, applied to the penetration of a cylindrical intruder near a vertical wall, as described in \cite{Diaz-Melian2020}. a

\begin{figure}
\includegraphics[width=200px]{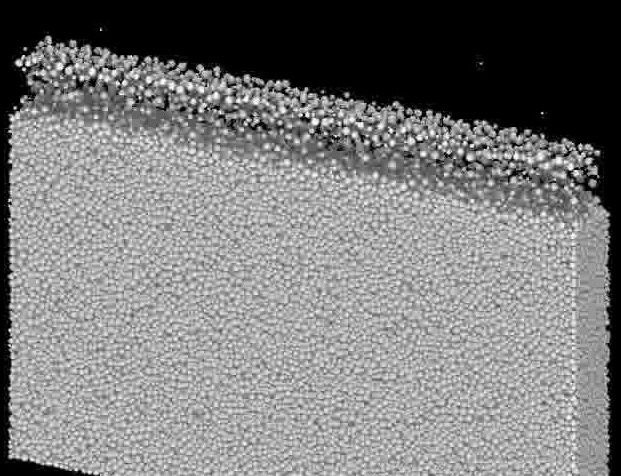}
\caption{Preparation of the granular bed in the DEM simulations. A group of particles is poured periodically at a certain height from the granular surface. }
\label{fig:prep}
\end{figure}

The numerical simulations were performed using LAMMPS \cite{plimpton2007lammps} reproducing the same experimental setup of \cite{Diaz-Melian2020}. The system consists in a granular bed of expanded polystirene spheres confined into a Hele-Shaw cell, where a large cylindrical intruder is released from the free granular surface by means of an electromagnetic device that minimizes initial spurious vibrations and torques. Before its release, the cylinder is gently touching the left vertical wall of the cell.

The simulations were divided into two stages: preparation of the granular bed and the release of the intruder. In the first, the granular bed is prepared by pouring batches of particles with radius following a uniform random distribution within the range $1 - 3.25$~mm and a fixed density of $14\times10^{-3}$~g/cc. Each pour generates particles at random positions within a limited space of the container, that moves up in the vertical direction as the container is filled (Fig. \ref{fig:prep}). The simulation ends when $10^5$ particles are deposited and the total kinetic energy drops below $10^{-7}$\,J.

\begin{figure}
\includegraphics[width=250px]{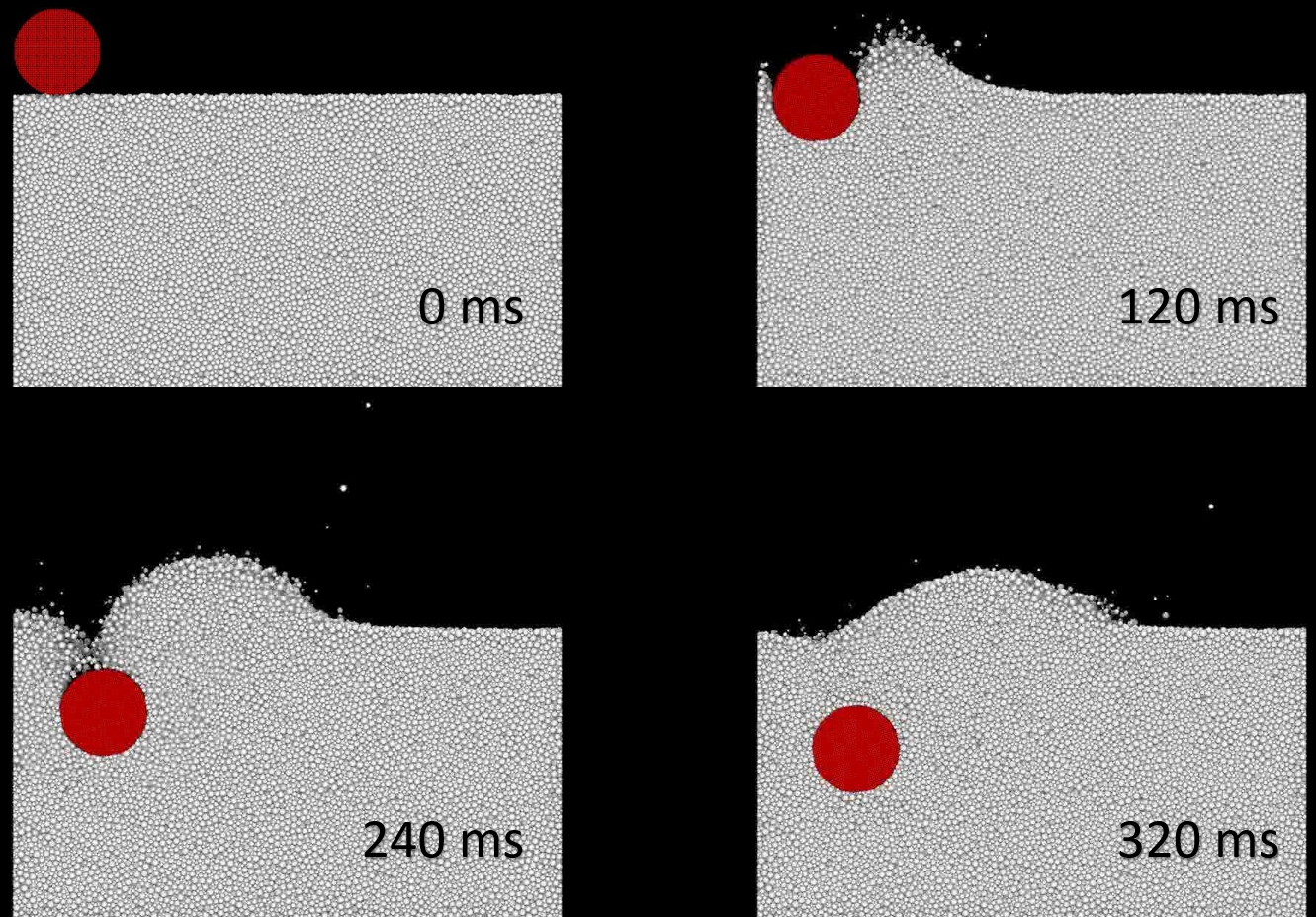}
\caption{Rendered images of the DEM simulations describing the motion of the intruder as it penetrates into the granular bed near the wall.}
\label{fig:snaps}
\end{figure}

After the first simulation ends, the intruder is generated as a $3.75\,$cm radius, $5.2\,$cm height and $0.255\,$kg mass cylinder, modeled as a many-particle-rigid-body, using a $1$~mm-spacing simple cubic lattice, formed by spherical particles with a diameter of $1$~mm. The intruder is released from the surface of the granular bed with zero initial velocity as shown at the upper left snapshot in Fig. \ref{fig:snaps}. The interaction between particles is ruled by a Hertzian contact model \cite{brilliantov1996model,silbert2001granular,zhang2005jamming} where the parameters are calculated based on the material properties: Young Modulus $E = 5\,GPa$, restitution coefficient $e = 0.1$, Poisson's ratio $\nu = 2/7$ and friction coefficient $\mu = 0.5$, as described in \cite{iliuta2003concept,yu2005improved}. Fig. \ref{fig:snaps} shows the typical motion of the intruder: as it penetrates it suffers a horizontal repulsion due to the effect of the wall. In this work, we will describe the impact of $\mu$, $E$ and $e$ in the penetration process, that is, the time evolution of the vertical and horizontal displacements of the intruder.

\begin{figure}
\includegraphics[width=220px]{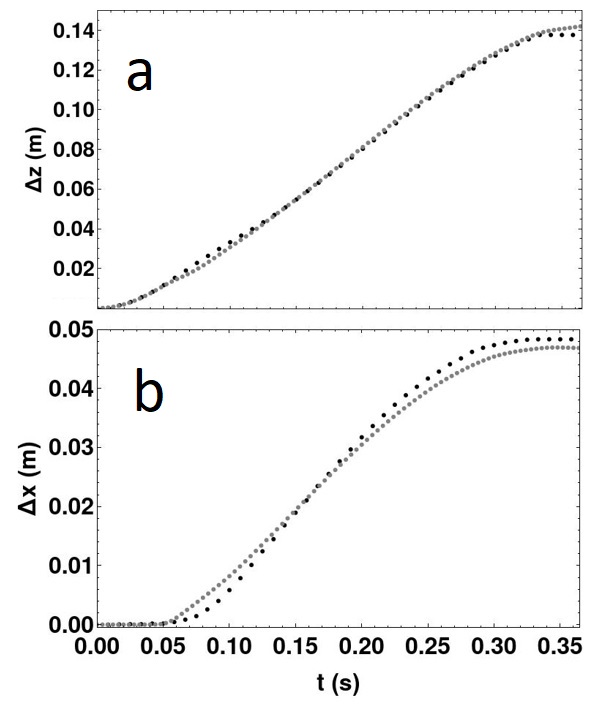}
\caption{DEM simulation vs experimental results. (a) Vertical penetration of the intruder (experiment:black, simulation:gray) as an average over 10 numerical simulations/experiments. (b) Horizontal displacement of the intruder (Experimental data was taken from \cite{Diaz-Melian2020}).}
\label{fig:XZfit}
\end{figure}

The first step is to check if the simulations are able to reproduce the experimental results. In Fig. \ref{fig:XZfit} we observe a good fit between experiment and simulation, within root-mean-square errors of 0.0554 and 0.0347 (normalized to the maximum value of each experimental curve), for the horizontal and vertical motion, respectively. This confirms the possibility of using DEM simulations to investigate further experimental situations by changing the material properties of the granular system. 

\begin{figure}
\includegraphics[width=220px]{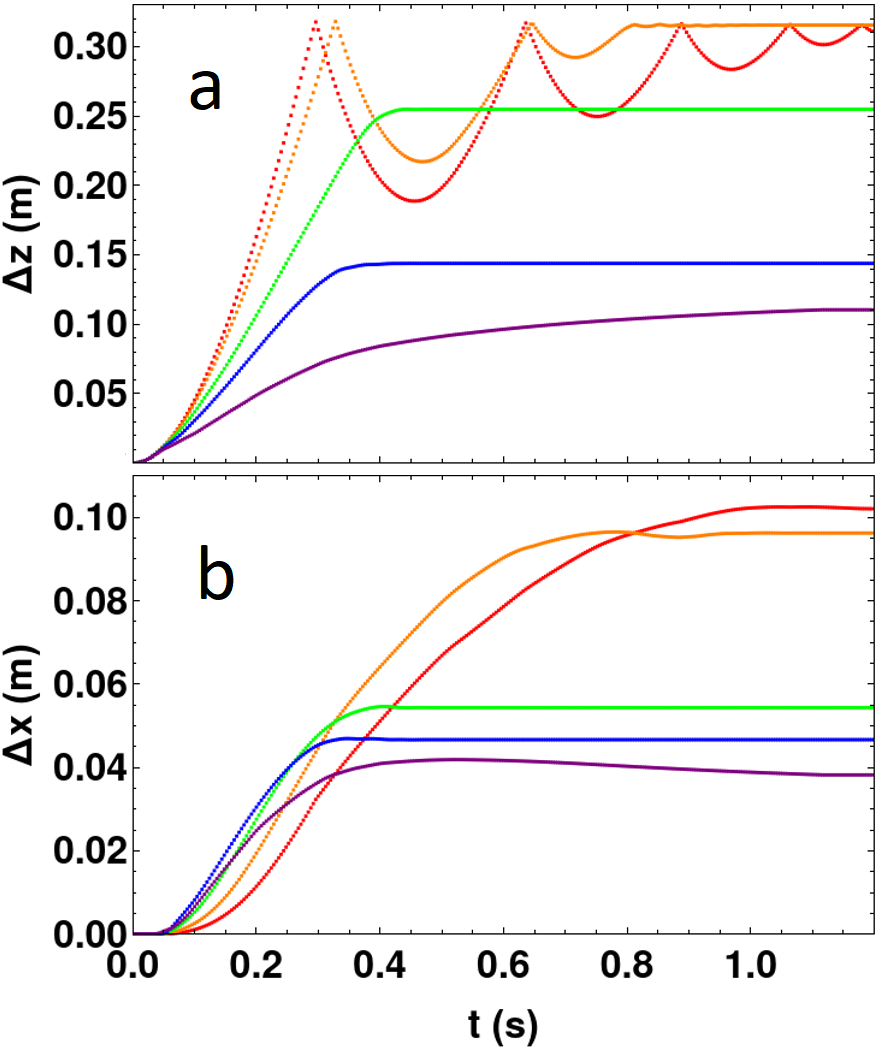}
\caption{Sensibility of the penetration process to changes in the friction coefficient. (a) Time evolution of the vertical motion of the intruder for $\mu = 0.05$ (red), $0.10$ (orange), $0.25$ (green), $0.50$ (blue), $0.90$ (purple). (b) Same as (a) describing the horizontal motion.}
\label{fig:fric}
\end{figure}

The friction coefficient $\mu$ plays an essential role in granular phenomena, as it causes the stress to be transmitted by relatively rigid, heavily stressed chains of particles forming a sparse network of contact forces known as ``force chains''. The bigger $\mu$ the more rigid and stable are the force chains. In Fig. \ref{fig:fric} we observe the effect of the value of $\mu$ on the motion of the cylindrical intruder. For low values (i.e. $\mu \le 0.1$) the intruder hits the bottom of the cell as the force chains are never ``stronger'' than the gravity acting on the intruder. For $\mu \ge 0.25$ the motion stops before reaching the floor; the higher the value of $\mu$ the lesser the vertical penetration.

Comparing Fig. \ref{fig:fric} (a) and (b) shows an interesting phenomenon: while the difference in the vertical penetration for $\mu = 0.25,\,0.50$ and $0.90$ is rather large, the difference in the horizontal repulsion is much smaller. This could mean that the initial force chains, i.e. those ``charged" after the initial vertical free-fall plunge of the intruder \cite{Diaz-Melian2020}, could be the ones with the largest component in the $x$-axis, repelling the intruder from the wall (this is consistent with the experiment and model predictions in \cite{Diaz-Melian2020}). Although higher values of $\mu$ lead to smaller penetration, it also increases the initial ``push" caused by the force chains formed between the intruder and the vertical wall, offsetting the difference in the horizontal displacement. 
 
\begin{figure}
\includegraphics[width=220px]{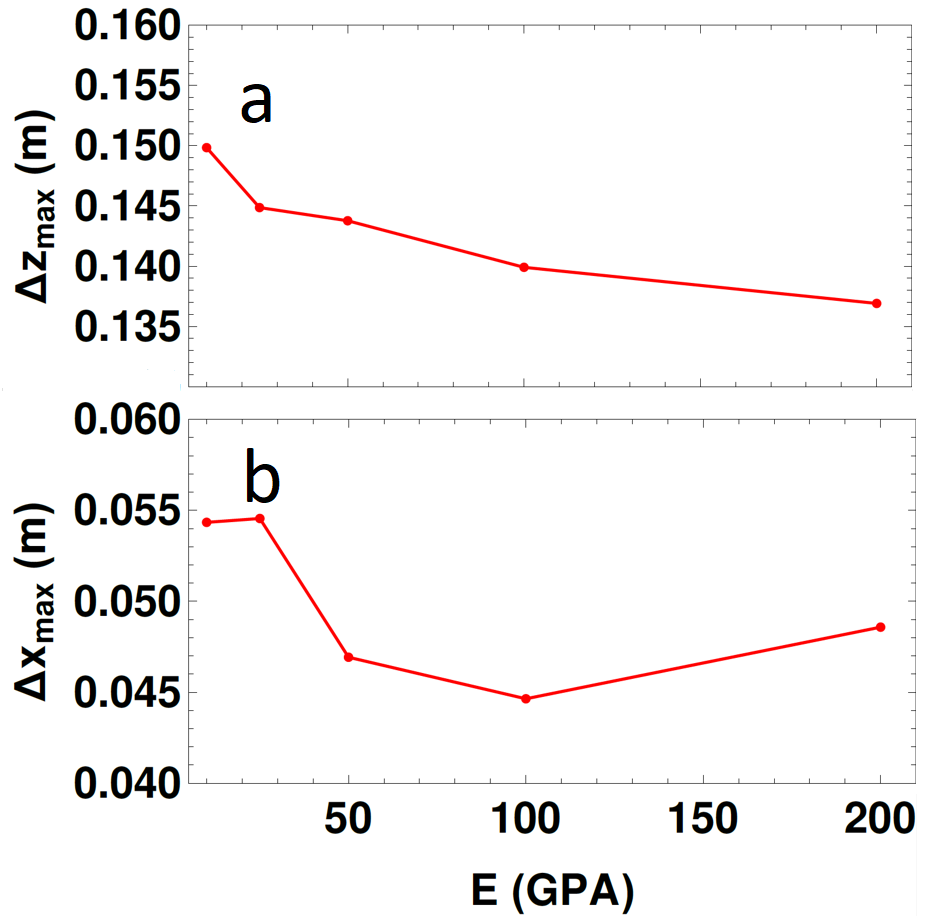}
\caption{Sensibility of the penetration dynamic to changes in the Young Modulus. (a) Final penetration depth as a function of $E$. (b) Maximum lateral displacement.}
\label{fig:young}
\end{figure}

The Young Modulus $E$ describes the relation between the compressive stress and the axial strain of a solid material. The greater $E$ the less deformation of the solid under the same force per unit area. Fig. \ref{fig:young} shows the sensibility of the studied granular system to changes in the Young Modulus of the grains. As can be observed, the impact, for the studied range of values, is less important than that caused by changes in friction. In Fig. \ref{fig:young} (b) a slight decrease in the vertical penetration of the intruder is observed as $E$ increases. This could be related to the increase in the resistance of the particles to deformation (i.e. stronger force chains). In any case, the relatively negligible effect of Young's modulus indicates that our granular system is, as a whole, more plastic than elastic: the building and destruction of force chains depends more on friction than on the bulk elastic properties of the grains.

The coefficient of restitution $e$ corresponds to the ratio between the final to initial relative speeds between two colliding particles, that is, the ratio of translational kinetic energy converted to other forms of energy (heat, deformation) during the process. 

\begin{figure}
\includegraphics[width=220px]{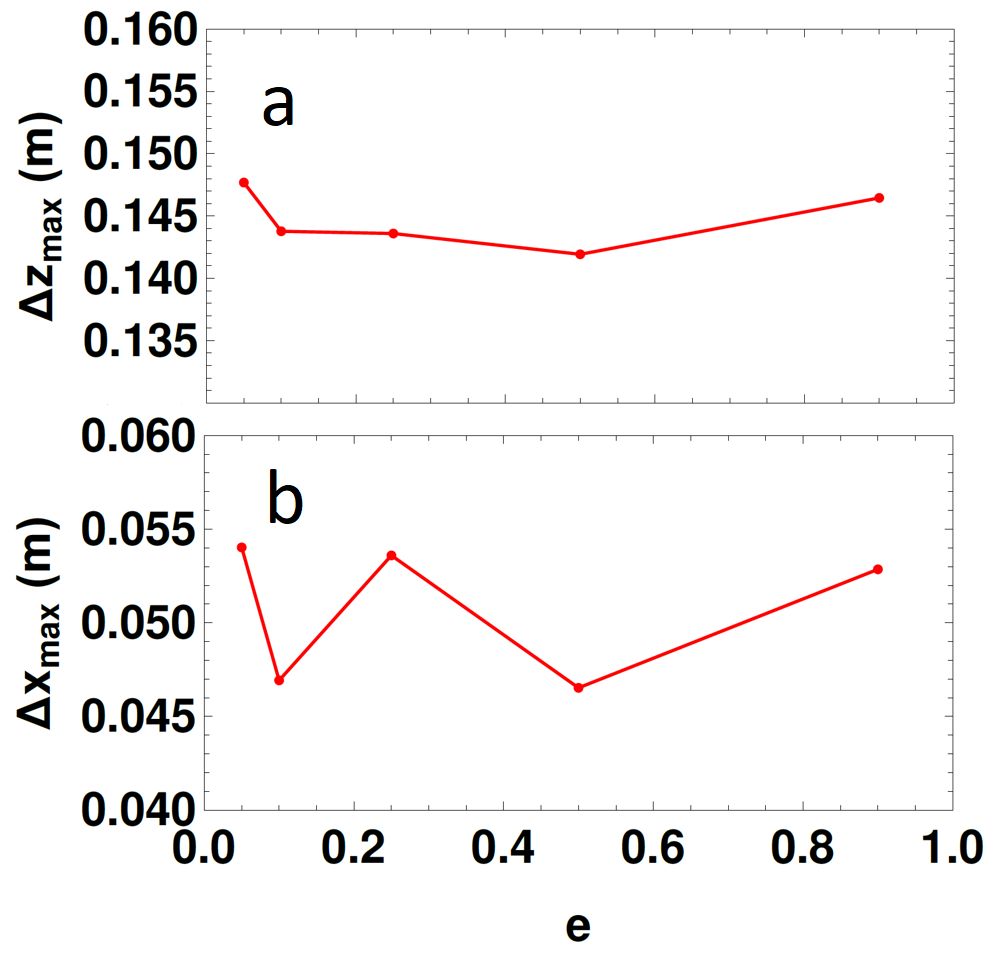}
\caption{Sensibility of the penetration process to changes in the restitution coefficient. (a) Final penetration depth  as a function of $e$. (b) Maximum lateral displacement.}
\label{fig:rest}
\end{figure}

In Fig. \ref{fig:rest} we observe that changes in $e$ cause negligible effect in the motion of the intruder. This could be related to two factors: one, a large part of the kinetic energy dissipates due to tangential dynamic (friction); and two, the large network of force chains allow for a sparse distribution of the force and dissipation of the kinetic energy of the intruder. The second reason is consistent with the relatively higher value of penetration observed for $e = 0.05$ in Fig. \ref{fig:rest} (a) which suggests that values of $e$ closer to zero offer less resistance to the motion of the intruder, as the only way to dissipate the kinetic energy would be through the friction between particles.

In summary, we have studied a granular system where a large and massive cylindrical intruder penetrates from the free granular surface with zero initial velocity near a wall, using DEM simulations. The sensibility of the system to different grain properties was analyzed, observing the great impact of the friction coefficient in the motion of the intruder. Higher values of $\mu$ allow the creation of stronger force chains that reduce the vertical penetration of the intruder, yet they also increase the initial horizontal repulsion form the lateral wall.

However, the influence of the Young modulus and the restitution coefficient of the grains have a negligible impact in the penetration process, suggesting that they play a less important role in the formation and strength of force chains.

\bibliography{biblio.bib}
\bibliographystyle{rcfbib}

\end{document}